# Charge states and magnetic ordering in $LaMnO_3/SrTiO_3$ superlattices

Woo Seok Choi[1*], D. W. Jeong[1], S. S. A. Seo[2**], Y. S. Lee[3], T. H. Kim[1], S. Y. Jang[1], H. N. Lee[2,a)], and K. Myung-Whun[4,b)]

[1]*ReCFI, Department of Physics and Astronomy, Seoul National University, Seoul 151-747, Korea*

[2]*Materials Science and Technology Division, Oak Ridge National Laboratory, Oak Ridge, Tennessee 37831, USA*

[3]*Department of Physics, Soongsil University, Seoul 156-743, Korea*

[4]*Department of Physics & IPIT, Chonbuk National University, Jeonju 561-756, Korea*

We investigated the magnetic and optical properties of $[(LaMnO_3)_n/(SrTiO_3)_8]_{20}$ ($n$ = 1, 2, and 8) superlattices grown by pulsed laser deposition. We found a weak ferromagnetic and semiconducting state developed in all superlattices. An analysis of the optical conductivity showed that the $LaMnO_3$ layers in the superlattices were slightly doped. The amount of doping was almost identical regardless of the $LaMnO_3$ layer thickness up to eight unit cells, suggesting that the effect is not limited to the interface. On the other hand, the magnetic ordering became less stable as the $LaMnO_3$ layer thickness decreased, probably due to a dimensional effect.

[*]Current address: Materials Science and Technology Division, Oak Ridge National Laboratory, Oak Ridge, Tennessee 37831, USA

[**]Current address: Department of Physics & Astronomy, University of Kentucky, Lexington, Kentucky 40506, USA

[a)]E-mail: hnlee@ornl.gov

[b)]E-mail: mwkim@chonbuk.ac.kr

## I. INTRODUCTION

A great deal of interest has been paid toward perovskite transition metal oxide heterostructures due to their versatile physical properties.[1] Among the novel properties, there have been efforts to control and utilize magnetic behaviors. For example, by exploiting manganese oxides as a constituent layer in the oxide heterostructures, the magnetoelectric coupling between magnetism and ferroelectricity, modification of the magnetic ordering, and charge transfer effect were observed.[2-5]

$LaMnO_3$-based layers have been widely adopted to employ magnetism into the oxide heterostructures.[5-8] One of the advantages of using $LaMnO_3$ is that its magnetic and electric phases can be modified diversely by a small amount of doping. Stoichiometric $LaMnO_3$ is an A-type antiferromagnet with a Néel temperature of ~140 K,[9] and a good insulator. The system can be doped by cautiously controlling its stoichiometry, and can become a ferromagnetic metal. On the other hand, such a doping effect can also be a major disadvantage in identifying the system. The delicate effect of doping requires the careful characterization of $LaMnO_3$, especially when it is used in a heterostructure.[8, 10-15] In particular, in oxide superlattices, few nanometers of thin $LaMnO_3$ layers are adjoined to other oxide layers. In such cases, the Mn valence near the interface may easily differ from its normal value of +3 owing to its multivalence nature.[8, 10-12]

Optical spectroscopy is an ideal tool for examining $LaMnO_3$-based oxide heterostructures. In contrast to other spectroscopic tools, of which the accessible range is usually limited to near the surface, optical spectroscopy employs low energy photons whose penetration depth is usually longer than the thickness of most oxide heterostructures. Therefore, one can look deeper into the heterostructure and characterize the buried manganese oxide layers. In addition, optical spectroscopy is a sensitive tool for picking up changes in the electronic structure of $LaMnO_3$ with different doping concentration.[16, 17] Therefore, when combined with electric and magnetic measurements, optical spectroscopy can characterize physical properties of $LaMnO_3$-based oxide heterostructures quite accurately.

In this paper, we grew different thicknesses of $LaMnO_3$ layers sandwiched between the $SrTiO_3$ layers, forming $[(LaMnO_3)_n/(SrTiO_3)_8)]_{20}$ ($n$ = 1, 2 and 8) superlattices and studied their magnetic and optical properties.

## II. EXPERIMENTS

High quality $[(LaMnO_3)_n/(SrTiO_3)_8)]_{20}$ superlattices on single unit cell stepped $SrTiO_3$ single-crystal substrates were grown epitaxially by pulsed laser deposition (PLD). The samples were grown at 700°C in an oxygen partial pressure of 10 mTorr, and the growth was monitored by observing the intensity oscillation of reflection high-energy electron diffraction specular spot. Three $[(LaMnO_3)_n/(SrTiO_3)_8)]_{20}$ superlattices were grown for this study, i.e. $n$ = 1, 2, and 8. The thickness of the $LaMnO_3$ layer was 1, 2, and 8 unit cells, while the thickness of the $SrTiO_3$ layer was fixed to 8 unit cells. The period of the superlattices was also fixed at 20. The superlattices were annealed at 900 °C for 1 hour in an oxygen-reducing atmosphere (4% H + 96% Ar), as PLD grown $LaMnO_3$ are usually of off-stoichiometry and hole doped due to the presence of excess oxygen ions.[16-20]

X-ray diffraction (XRD) and reciprocal space mapping of the annealed superlattices were measured at the synchrotron radiation source at the 10C1 beam-line of the Pohang Accelerator Laboratory. The in-plane transmittance ($T(\omega)$) and reflectance ($R(\omega)$) spectra were measured in the near-normal geometry. A Fourier-transform-type infrared spectrometer was used to take the spectra between 0.07 and 1.5 eV. A grating-type spectrophotometer was employed to measure the spectra between 0.4 and 5.9 eV. The real part of the in-plane optical conductivity $\sigma_1(\omega)$ at 0.3 – 3.2 eV (where the $SrTiO_3$ substrate was transparent) was obtained using a numerical iteration intensity-transfer-matrix method. The temperature- ($T$-) and magnetic field- ($H$-) dependent magnetization ($M$, zero-field cooled, while cooling) was measured using a SQUID magnetometer. $H$ was applied along the $a$-axis.

## III. RESULTS AND DISCUSSION

### A. Structural characterization of $LaMnO_3/SrTiO_3$ superlattices

Figure 1(a) shows a XRD $\theta$-$2\theta$ scan of the $[(LaMnO_3)_8/(SrTiO_3)_8)]_{20}$ superlattice. The clear superlattice satellite peaks along with thickness fringe reflections indicate well-defined surface and interfaces. Sixteen satellite peaks were identified between the main peaks and the thickness of one superlattice period was 6.265 nm, almost 16 times of the perovskite unit cell thickness, as intended for $[(LaMnO_3)_8/(SrTiO_3)_8)]_{20}$ superlattice. No other Bragg peaks were observed, suggesting that the superlattice has good crystallinity without secondary phases. From the rocking curve we obtained the full-width-half-maximum values of the peaks as 0.028, 0.030 and 0.053 for $n$ = 1, 2, and 8 $[(LaMnO_3)_n(SrTiO_3)_8]_{20}$ superlattices respectively. Note that the values are comparable to that of single crystal $SrTiO_3$ substrates, indicating good crystallinity of our superlattices. It is also worthwhile to mention the asymmetric shape around the substrate 002 peak in the XRD $\theta$-2$\theta$ scan. The intensity below the 002 peak is larger than that above the peak. This suggests that the $LaMnO_3$ layers within the superlattice were compressively strained, leading to a slight elongation of the *c*-axis lattice constant and tetragonal-like structure.

Figure 1(b) shows an x-ray reciprocal space map of the $[(LaMnO_3)_8/(SrTiO_3)_8)]_{20}$ superlattice around the $SrTiO_3$ 103 Bragg peak. The superlattice peaks were readily apparent. All the superlattice peaks are on the same *h* value line with the substrate peak, indicating that the superlattices are under coherent compressive strain without in-plane lattice relaxation. The other $[(LaMnO_3)_n/(SrTiO_3)_8)]_{20}$ superlattices with $n$ = 1 and 2 also showed similar XRD results ensuring the high quality of these samples.

Figure 2 shows a schematic cross-section of ideal $[(LaMnO_3)_n/(SrTiO_3)_8)]_{20}$ superlattices. The plane of A-site ions (La and Sr) is half a unit cell shifted toward the normal of the cross-section from the B-site ions (Mn and Ti) and O-site ions (oxygen) in the perovskite $ABO_3$ structure. Since we used $TiO_2$ layer terminated $SrTiO_3$ substrate, the stacking sequence of $LaMnO_3$ on top of $TiO_2$ should start from LaO layer and be terminated with $MnO_2$ layer. That of $SrTiO_3$ on top of the $LaMnO_3$ should start from SrO layer, which makes the interface chemically asymmetric.. For the $n$ = 1 superlattice, the stacking sequence is (SrO)-($TiO_2$)-(*LaO*)-($MnO_2$)-(*SrO*)-($TiO_2$)-(SrO). Therefore, the $MnO_2$ layers in the $n$ = 1 superlattice lie in a different chemical environment compared to bulk $LaMnO_3$. We call this $MnO_2$ layer in an asymmetric chemical environment as the 'interfacial layer'. On the other hand, while $n$ = 2 and 8 superlattices possess interfacial $MnO_2$ layers, they also have the chemically symmetric $MnO_2$ layers resembling those of bulk $LaMnO_3$. We call the symmetric $MnO_2$ layers as the 'inner layers'.

### B. Magnetic properties of $LaMnO_3/SrTiO_3$ superlattices

Figure 3(a) shows the temperature dependent magnetization, $M(T)$, of the $[(LaMnO_3)_n/(SrTiO_3)_8)]_{20}$ superlattices. The $M(T)$ curves of all the superlattices showed increasing behavior with decreasing $T$, but the shape of the curves was quite different depending on the $n$ values. The $M(T)$ curve of the $n$ = 1 superlattice shows a slow increase and a slight change in slope below ~25 K as the temperature decreases. The $M(T)$ curve of the $n$ = 2 superlattice also shows a similar increase with decreasing temperature but a change in slope occurs at higher temperatures (~70 K) compared to the $n$ = 1 superlattice. The $M(T)$ curve of the $n$ = 8 superlattice showed an abrupt increase below ~120 K with the steepest slope, which is reminiscent of the $M(T)$ curves of the ferromagnetic metallic perovskite manganese oxides. The ferromagnetic ordering temperature increased with increasing number of $MnO_2$ layers. Figure 3(a) also shows the $M(T)$ curves of $LaMnO_3$ thin films for comparison. It shows a weaker temperature dependence with only a slight enhancement below ~130 K, which is similar to that of the bulk $LaMnO_3$.[21] The $M(T)$ curves of the $[(LaMnO_3)_n/(SrTiO_3)_8)]_{20}$ superlattices at low temperatures were mostly laid above the $LaMnO_3$ thin films.

Figure 3(b) shows the isothermal $M(H)$ curves at 10 K. The $M(H)$ curves of the $LaMnO_3$ thin films are also shown for comparison. While the saturated magnetic moment of $LaMnO_3$ thin film were ~0.5 $\mu_B$/Mn, close to that of the stoichiometric $LaMnO_3$ crystal,[21] the saturated magnetic moment of the superlattices were quite larger. Although the low magnetic field region is different, the saturated magnetic moment of the $[(LaMnO_3)_n/(SrTiO_3)_8)]_{20}$ superlattice is commonly 1.3 ± 0.3 $\mu_B$/Mn, which is almost independent of the $n$ value.

The large magnetic moment might come from Ti ions which could form a ferromagnetic state in the interfacial $TiO_2$ layer. However, the magnetic contribution from the Ti ions in Sr-doped $LaTiO_3$ or in $LaTiO_{3+\delta}$ is much weaker than that from Mn ions.[22] Therefore, the observed magnetization in $M(T)$ curve should mostly be due to the Mn spins. On the other hand, Mn spins in the interfacial $MnO_2$ layer might be considered as the origin of the ferromagnetic state in the $[(LaMnO_3)_n/(SrTiO_3)_8)]_{20}$ superlattices. However, if the ferromagnetic state is solely due to the interfacial layer, there should be antiferromagnetic or paramagnetic $LaMnO_3$ layers remaining in the $n$ = 2 and $n$ = 8 superlattices. In this case, the saturated magnetic moment should decrease systematically with increasing $n$ because the magnetization yields volume averaged information on the magnetic moment. The almost $n$ independent saturated magnetic moment suggests that the magnetic state of the local Mn spins in the inner layer is similar to that in the interfacial layers.

The thickness independent saturated magnetic moment led us to compare the thin $LaMnO_3$ layers with their bulk state. A similar weak ferromagnetic state appears in the bulk $LaMnO_{3+\delta}$ crystal for $\delta = 0.025$ and for $\delta = 0.15$.[21] Chemical analysis showed that the bulk states were due to 5% ($\delta = 0.025$) and 30% ($\delta = 0.15$) of $Mn^{4+}$, respectively. According to neutron scattering analysis, for $\delta = 0.025$, the weak ferromagnetic state is the result of the coexistence of ferro and antiferromagnetic regions at low temperatures or the existence of a low-temperature canted antiferromagnetic structure. For $\delta = 0.15$, the weak ferromagnetic ordering is a result of a disordered cluster-glass state.[21] The magnetic structure of $[(LaMnO_3)_n/(SrTiO_3)_8)]_{20}$ superlattices can be associated with that of bulk crystal, once the doping concentration in the superlattices can be estimated.

**C. Optical properties of $LaMnO_3/SrTiO_3$ superlattices**

To obtain the doping concentration, we performed optical spectroscopy with the superlattices. Figure 4(a) shows $T(\omega)$ and the $R(\omega)$ of the $[(LaMnO_3)_n/(SrTiO_3)_8)]_{20}$ superlattices at room temperature. The abrupt change in $T(\omega)$ and $R(\omega)$ below 0.2 eV and above 3.2 eV are due to the strong absorption of the $SrTiO_3$ substrate. Below 3.2 eV, $T(\omega)$ of the superlattices decreases with increasing $LaMnO_3$ thickness. $SrTiO_3$ does not have any spectral features between 0.2 eV to 3.2 eV, whereas $LaMnO_3$ has an absorption peak structure centered at approximately 2 eV.[23] Therefore, the features between 0.2 eV and 3.2 eV in the superlattices originate mostly from the $LaMnO_3$ layers within the superlattices. $T(\omega)$ and $R(\omega)$ below 0.1 eV of the superlattices are almost identical to those of the $SrTiO_3$ substrate, which indicates that the charge carriers do not form a good metallic state. Therefore, the doping concentration can be estimated quite accurately by analyzing the optical spectra above ~0.2 eV.

Figure 4(b) shows $\sigma_1(\omega)$ of the $[(LaMnO_3)_n/(SrTiO_3)_8)]_{20}$ superlattices obtained from $T(\omega)$ and $R(\omega)$. The $\sigma_1(\omega)$ of all superlattices show an insulating optical gap of ~0.5 eV. The optical gap decreases with increasing $n$. The inset in Fig. 4(b) presents the dc transport measurement.[24] The resistivity, $\rho$, decreases with increasing $n$, consistent with the optical gap, whereas the overall insulating or semiconducting behavior is conserved and the curvature of $\rho$ reflecting the band gap is weakly dependent on $n$. Therefore the major cause of the optical gap decrease may not be the change in the band structure but probably be some impurity states near the band edge.

$\sigma_1(\omega)$ is the effective response of both $LaMnO_3$ and $SrTiO_3$ layers in the superlattices, so the spectral weight of the $LaMnO_3$ layers in the superlattices are underestimated. To properly estimate the contribution of the $LaMnO_3$ layers to $\sigma_1(\omega)$, $\sigma_1(\omega)$ of $LaMnO_3$ layer only, i.e. $\sigma_{1,\mathrm{LMO}}(\omega)$, was deduced by performing a two-dimensional effective medium approximation: [25]

$$\tilde{\varepsilon}_{\mathrm{SL}}^{eff} = \frac{\tilde{\varepsilon}_{\mathrm{LMO}} d_{\mathrm{LMO}} + \tilde{\varepsilon}_{\mathrm{STO}} d_{\mathrm{STO}}}{d_{\mathrm{LMO}} + d_{\mathrm{STO}}}, \qquad (1)$$

where $\tilde{\varepsilon}$ and $d$ correspond to the in-plane complex dielectric constants ($\tilde{\varepsilon} = \varepsilon_1 + i\varepsilon_2$, $\sigma_1(\omega) = \omega\varepsilon_2(\omega)/4\pi$) and the thickness of each layer, respectively. Here, $\tilde{\varepsilon}_{\mathrm{SL}}^{eff}$, $d_{\mathrm{LMO}}$ and $d_{\mathrm{STO}}$ are known, and $\tilde{\varepsilon}_{\mathrm{STO}}$ is measured for a $SrTiO_3$ single crystal annealed under the same conditions. Consequently, Eq. (1) yields the complex dielectric function of $LaMnO_3$ layers only, i.e. $\tilde{\varepsilon}_{\mathrm{LMO}}(\omega)$, thus, $\sigma_{1,\mathrm{LMO}}(\omega)$. Figure 4(c) shows the $\sigma_{1,\mathrm{LMO}}(\omega)$ extracted from the calculation.

Surprisingly, $\sigma_{1,\mathrm{LMO}}(\omega)$ was almost identical for all $[(LaMnO_3)_n/(SrTiO_3)_8)]_{20}$ superlattices studied in this work. For the $n$ = 1 superlattice, $\sigma_{1,\mathrm{LMO}}(\omega)$ should be due solely to the interfacial $LaMnO_3$ layers. On the other hand, for the $n$ = 2 and 8 superlattices, the $\sigma_{1,\mathrm{LMO}}(\omega)$ is an averaged response of the interfacial and inner $LaMnO_3$ layers. To understand the $n$ independence of $\sigma_{1,\mathrm{LMO}}(\omega)$, one might assume that only the interfacial $LaMnO_3$ layer is optically active and the inner $LaMnO_3$ layer does not contribute to $\sigma_{1,\mathrm{LMO}}(\omega)$. However, $LaMnO_3$ shows strong absorption at ~2 eV in most environments.[16, 20] Therefore, the spectral feature of the superlattices results from an averaged response of both interfacial and inner $LaMnO_3$ layers.

Figure 4(c) also presents $\sigma_1(\omega)$ of an undoped and 10 % doped $LaMnO_3$ film for comparison. $\sigma_1(\omega)$ of the undoped film shows an optical gap of ~1 eV. As observed in $R_{1-x}Sr_xTiO_{3+\delta}$ or $R_{1-x}Ca_xTiO_{3+\delta}$ ($R$ = rare earth), a small change in the doping concentration can modify $\sigma_1(\omega)$ below the optical gap quite significantly in most transition metal oxides near the filling-controlled metal-insulator-transition boundary.[26] The $\sigma_{1,\mathrm{LMO}}(\omega)$ of the $n$ = 1 superlattice shows a spectral weight below the gap of the undoped $LaMnO_3$ thin film, indicating the doping concentration in the interfacial $MnO_2$ layers is changed. However, there was no further significant modification of $\sigma_{1,\mathrm{LMO}}(\omega)$ due to the inner $MnO_2$ layers, as shown for the $n$ = 2 and 8 superlattices.

The observation demonstrates that the average doping concentration of the superlattice changes abruptly because of the interfacial layers, but changes only slightly due to the addition of the inner layers. The charges at the interfacial $LaMnO_3$ layer are most probably caused by the charge transfer from the adjacent $SrTiO_3$ layers for the $n$ = 1 superlattice. If the doping concentration change is limited to the interfacial layer only and the inner $LaMnO_3$ layers remain highly insulating as the undoped $LaMnO_3$ film, then the spectral weight below ~ 1 eV, the averaged response should become reduced as the $n$-value increases. However the spectral weights of the $n$ = 2 and 8 superlattices slightly increase, which indicate that the doping concentration of the inner $LaMnO_3$ layers is similar to that of interfacial layer. The origin of the extra charges at the inner layer is not clearly identified. There are a few candidates as discussed later.

It would be informative to compare the $\sigma_{1,\mathrm{LMO}}(\omega)$ of the superlattices with the $\sigma_1(\omega)$ of the homogenous thin films in order to estimate the doping concentration. $\sigma_1(\omega)$ of the 10% doped $LaMnO_3$ film shows some spectral weight below ~1 eV. Similarly, $\sigma_{1,\mathrm{LMO}}(\omega)$ of the $[(LaMnO_3)_n/(SrTiO_3)_8)]_{20}$ superlattices also show some spectral weight below 1 eV. However, the larger optical gap and smaller spectral weight around 1 eV clearly indicate that the doping concentration of the superlattices is smaller than that of the 10% doped thin film. Therefore, the doping level of the superlattice should be <10%, and similar to the bulk state of $\delta = 0.025$. The doped holes are responsible for the semiconducting behavior and ferromagnetic state. Without the holes, antiferromagnetic superexchange would dominate the magnetic interaction. With the holes, the ferromagnetic double exchange interaction plays a role. Thus, the observed weak ferromagnetic state in the superlattices is most likely due to the coexistence of the ferro- and antiferromagnetic interactions caused by the holes. However it should be noted that similar ferromagnetic states were also found in some manganites where the $Mn^{3+}$ ions were substituted by the isovalent Sc or Ga ions,[27, 28] which means that the existence of holes due to $Mn^{4+}$ ion is not the sufficient condition for the ferromagnetic state. Clear identification of the ferromagnetic state origin should be done in further studies.

**D. Origin of the doping in $LaMnO_3/SrTiO_3$ superlattices**

The hole doping might be caused by intrinsic origins. For an example, $La_{1/3}Sr_{2/3}MnO_3$ thin films grown on various single crystal substrates show very different magnetic and electric phases depending on the lattice strain even without chemical doping.[29] Similar phenomena could occur in the $LaMnO_3$ layers due to the strain from the neighboring $SrTiO_3$ layers. However, the doping does not appear to be associated with the strain effect because the $LaMnO_3$ thin film is also under the strain effect but is antiferromagnetic and highly insulating.

A more likely cause is the chemical structure around the Mn ions of the superlattices, which is quite different from that of the bulk $LaMnO_3$ crystal. A theoretical calculation suggested charge transfer between the $LaMnO_3$ and $SrMnO_3$ layers over a few unit cells in $LaMnO_3/SrMnO_3$ superlattices.[30] Transmission electron microscopy also showed that charge transfer can occur from the $LaMnO_3$ layers to $SrTiO_3$ layers near the symmetric interfaces of $LaMnO_3/SrTiO_3$ superlattices.[8] A similar charge transfer or charge smearing effect might cause hole doping in our $[(LaMnO_3)_n/(SrTiO_3)_8)]_{20}$ superlattices.

Meanwhile, structural defects and ion vacancies can be also responsible for doping. Although the superlattices were annealed, a small portion of the excess oxygen might remain in the superlattices. It might not be easy for thin $LaMnO_3$ layers to lose sufficient excess oxygen to retrieve its stoichiometric phase by thermal annealing in an environment where the $SrTiO_3$ layers sandwich the $LaMnO_3$ layers. Furthermore, the $LaMnO_3$ layers in the superlattices are fully strained and structurally clamped to the $SrTiO_3$ layers, which may worsen the situation. Therefore, the role of the remnant excess oxygen ions and/or the metal-ion vacancies cannot be excluded completely for the superlattices, even after annealing. This might be a major origin of the small difference between $n$ = 1 and $n$ = 2 or 8 superlattices. Another minor cause can be cation intermixing at the interface between $LaMnO_3$ and $SrTiO_3$. However, this is less probable because the interfaces and surfaces of the superlattices are well-defined according to the XRD measurements.

**E. Dimensional effect in $LaMnO_3$/$SrTiO_3$ superlattices**

Optical data demonstrates that localized doped holes contribute to form a weak ferromagnetic state in the $LaMnO_3$ layers, and the doping concentrations of all superlattices are similar. Therefore, if those $LaMnO_3$ layers are in the doped bulk $LaMnO_3$ state, their magnetic properties should also be similar to each other. However, it is quite intriguing that the detailed magnetic properties are quite different. As discussed above, the magnetic ordering temperature systematically decreases with decreasing $n$ in the $[(LaMnO_3)_n/(SrTiO_3)_8)]_{20}$ superlattices. In addition, the remnant $M$ value in the $M(H)$ curve decreases with decreasing $n$ to 1 or 2. The remnant $M$ value of the $n$ = 8 superlattice is 0.84 $\mu_B$/Mn, whereas the $n$ = 1 and 2 superlattices show a drastically decreased value of ~0.17 and ~0.09 $\mu_B$/Mn, respectively. These results indicate that magnetic ordering is affected by the effective dimension of the $LaMnO_3$ layer in the superlattices. Owing to the reduced dimension in the superlattice with thin $LaMnO_3$ layer, the long range magnetic ordering appears to be weakened.[31] The increased optical gap and more insulating behavior with decreasing $n$ value might also suggest that the magnetic ordering is disturbed at lower dimension. Further experimental investigations of the spin state will be needed to elucidate more details on the magnetic ordering.

Finally, it is interesting that three small peak structures at ~2 eV are commonly observed for $n$ = 1, 2, and 8 superlattices and their spectral shapes are also almost identical. The origin of the peak structure is not completely understood.[32, 33] However, for a more comprehensive understanding of the origin of the small peak structure, it should be noted that these small peaks appear also in the doped single interfacial $MnO_2$ layer lying between a SrO layer and LaO layer.

## IV. SUMMARY

We examined the magnetic, optical, and electrical properties of $[(LaMnO_3)_n/(SrTiO_3)_8)]_{20}$ ($n$ = 1, 2, and 8) superlattices. Ferromagnetic states were formed in the superlattices. and the $LaMnO_3$ layers in the superlattice were found to be slightly hole doped. The doping concentration was estimated to be <10%, consistently for the different thickness of $LaMnO_3$ layers within the superlattices. On the other hand, the magnetic ordering changed with the thickness of the $LaMnO_3$ layer and therefore, different from that of doped bulk $LaMnO_3$. The effect of the reduced dimensions was suggested to be the origin.

## ACKNOWLEDGEMENTS

The authors are grateful for valuable discussion with J.-S. Chung and H. Christen. This research was supported by the National Research Foundation of Korea (NRF) grants funded by the Korean government (MEST) (Grants No. 2009-0080567, 2010-0020416, and 2010-0014488). The experiments at Pohang Accelerator Laboratory were supported in part by MEST and Pohang University of Science and Technology. The work at Oak Ridge National Laboratory was sponsored by the Materials Sciences and Engineering Division, U.S. Department of Energy.

FIGURE CAPTIONS

FIG. 1. (Color online) (a) X-ray diffraction $\theta$-$2\theta$ scan around the $SrTiO_3$ 002 (*) reflection and (b) X-ray reciprocal space map around the substrate $SrTiO_3$ 103 Bragg peak recorded from a $[(LaMnO_3)_n/(SrTiO_3)_8)]_{20}$ ($n$ = 8) superlattice. Well-defined superlattice satellite peaks are clearly observed, confirming atomic scale control of the layer thickness and intended periodicity of the superlattice.

FIG. 2. (Color online) Schematic cross-section of an ideal $[(LaMnO_3)_n/(SrTiO_3)_8)]_{20}$ (a) $n$ = 1 and (b) $n$ = 2 superlattices. The plane of La and Sr ions are half a unit cell shifted toward the normal of the cross-section from the Ti, Mn and O ions.

FIG. 3. (Color online) (a) $M(T)$ and (b) $M(H)$ curves for $[(LaMnO_3)_n/(SrTiO_3)_8)]_{20}$ superlattices. $n$ = 1, 2, and 8 are shown as red solid, green dashed, and blue dotted lines, respectively. For comparison, similar curve for $LaMnO_3$ film is also shown in black thin solid line.[16]

FIG. 4. (Color online) (a) $T(\omega)$ (thick lines) and $R(\omega)$ (thin lines) for $[(LaMnO_3)_n/(SrTiO_3)_8)]_{20}$ superlattices. $n$ = 1, 2, and 8 are shown as red solid, green dashed, and blue dotted lines, respectively. (b) $\sigma_1(\omega)$ for the $[(LaMnO_3)_n/(SrTiO_3)_8)]_{20}$ superlattices obtained from ITMM method. The inset shows $\rho(T)$ curves. (c) $\sigma_{1,LMO}(\omega)$ of the $LaMnO_3$ layers of the superlattices, extracted from a numerical analyses (see the text). For comparison, $\sigma_1(\omega)$ for the stoichiometric and hole doped $LaMnO_3$ films are also shown in black thin solid and gray dash-dotted lines, respectively.[16]

REFERENCES

[1] S. B. Ogale, *Thin Films and Heterostructures for Oxide Electronics* (Springer, New York, 2005).

[2] V. Garcia, et al., Science **327**, 1106 (2010).

[3] C. A. F. Vaz, J. Hoffman, Y. Segal, J. W. Reiner, R. D. Grober, Z. Zhang, C. H. Ahn, and F. J. Walker, Phys. Rev. Lett. **104**, 127202 (2010).

[4] N. Kida, H. Yamada, H. Sato, T. Arima, M. Kawasaki, H. Akoh, and Y. Tokura, Phys. Rev. Lett. **99**, 197404 (2007).

[5] S. J. May, et al., Nature Mater. **8**, 892 (2009).

[6] B. R. K. Nanda and S. Satpathy, Phys. Rev. B **81**, 224408 (2010).

[7] H. Yamada, M. Kawasaki, T. Lottermoser, T. Arima, and Y. Tokura, Appl. Phys. Lett. **89**, 052506 (2006).

[8] J. Garcia-Barriocanal, et al., Adv. Mater. (Weinheim, Ger.) **22**, 627 (2010).

[9] E. O. Wollan and W. C. Koehler, Phys. Rev. **100**, 545 (1955).

[10] J. Garcia-Barriocanal, et al., Nat Commun **1**, 82 (2010).

[11] A. B. Shah, et al., Adv. Mater. (Weinheim, Ger.) **22**, 1156 (2010).

[12] H. M. Christen, D. H. Kim, and C. M. Rouleau, Appl. Phys. A: Mater. Sci. Process. **93**, 807 (2008).

[13] S. Dong, R. Yu, S. Yunoki, G. Alvarez, J. M. Liu, and E. Dagotto, Phys. Rev. B **78**, 201102 (2008).

[14] C. Aruta, et al., Phys. Rev. B **80**, 140405 (2009).

[15] X. Zhai, C. S. Mohapatra, A. B. Shah, J.-M. Zuo, and J. N. Eckstein, Adv. Mater. (Weinheim, Ger.) **22**, 1136 (2010).

[16] W. S. Choi, et al., J. Phys. D: Appl. Phys. **42**, 165401 (2009).

[17] W. S. Choi, D. W. Jeong, S. Y. Jang, Z. Marton, S. S. A. Seo, H. N. Lee, and Y. S. Lee, J. Korean Phys. Soc., unpublished (2010).

[18] H. S. Kim and H. M. Christen, J. Phys.: Condens. Matter **22**, 146007 (2010).

[19] W. Prellier, M. Rajeswari, T. Venkatesan, and R. L. Greene, Appl. Phys. Lett. **75**, 1446 (1999).

[20] Z. Marton, S. S. A. Seo, T. Egami, and H. N. Lee, J. Cryst. Growth **312**, 2923 (2010).

[21] C. Ritter, M. R. Ibarra, J. M. De Teresa, P. A. Algarabel, C. Marquina, J. Blasco, J. Garc, S. Oseroff, and S. W. Cheong, Phys. Rev. B **56**, 8902 (1997).

[22] Y. Okada, T. Arima, Y. Tokura, C. Murayama, and N. Mori, Phys. Rev. B **48**, 9677 (1993).

[23] N. N. Kovaleva, A. V. Boris, C. Bernhard, A. Kulakov, A. Pimenov, A. M. Balbashov,

G. Khaliullin, and B. Keimer, Phys. Rev. Lett. **93**, 147204 (2004).

[24]

[25] W. S. Choi, H. Ohta, S. J. Moon, Y. S. Lee, and T. W. Noh, Phys. Rev. B **82**, 024301 (2010).

[26] T. Katsufuji, Y. Okimoto, and Y. Tokura, Phys. Rev. Lett. **75**, 3497 (1995).

[27] J. S. Zhou and J. B. Goodenough, Phys. Rev. B **77**, 172409 (2008).

[28] J. B. Goodenough, R. I. Dass, and J. Zhou, Solid State Sciences **4**, 297 (2002).

[29] Y. Tokura and N. Nagaosa, Science **288**, 462 (2000).

[30] C. Lin, S. Okamoto, and A. J. Millis, Phys. Rev. B **73**, 041104 (2006).

[31] F. Bloch, Z. Phys. **61**, 206 (1930).

[32] A. S. Moskvin, A. A. Makhnev, L. V. Nomerovannaya, N. N. Loshkareva, and A. M. Balbashov, Phys. Rev. B **82**, 035106 (2010).

[33] M. W. Kim, P. Murugavel, S. Parashar, J. S. Lee, and T. W. Noh, New J. Phys. **6**, 156 (2004).

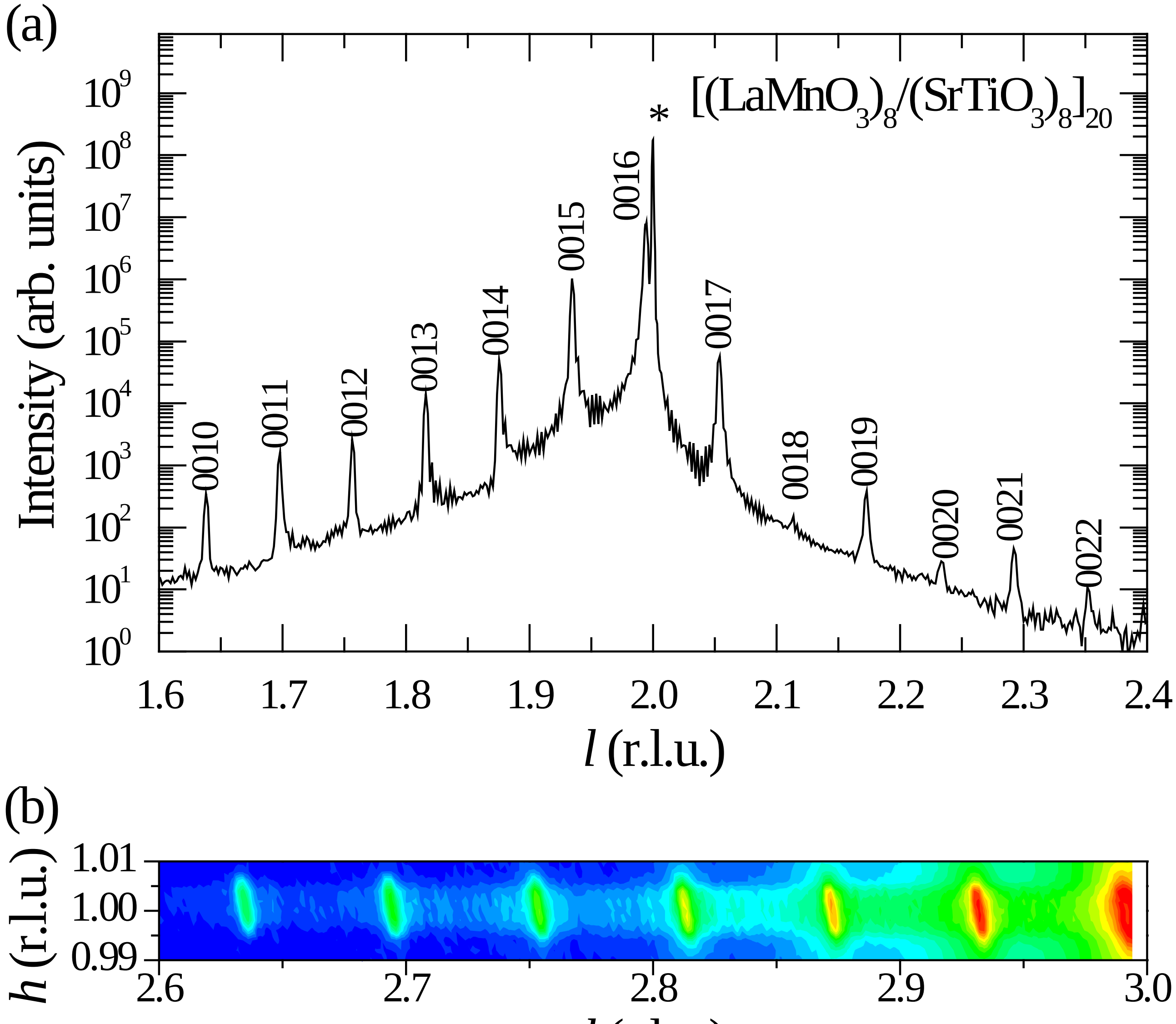

Fig. 1.
Choi *et al*.

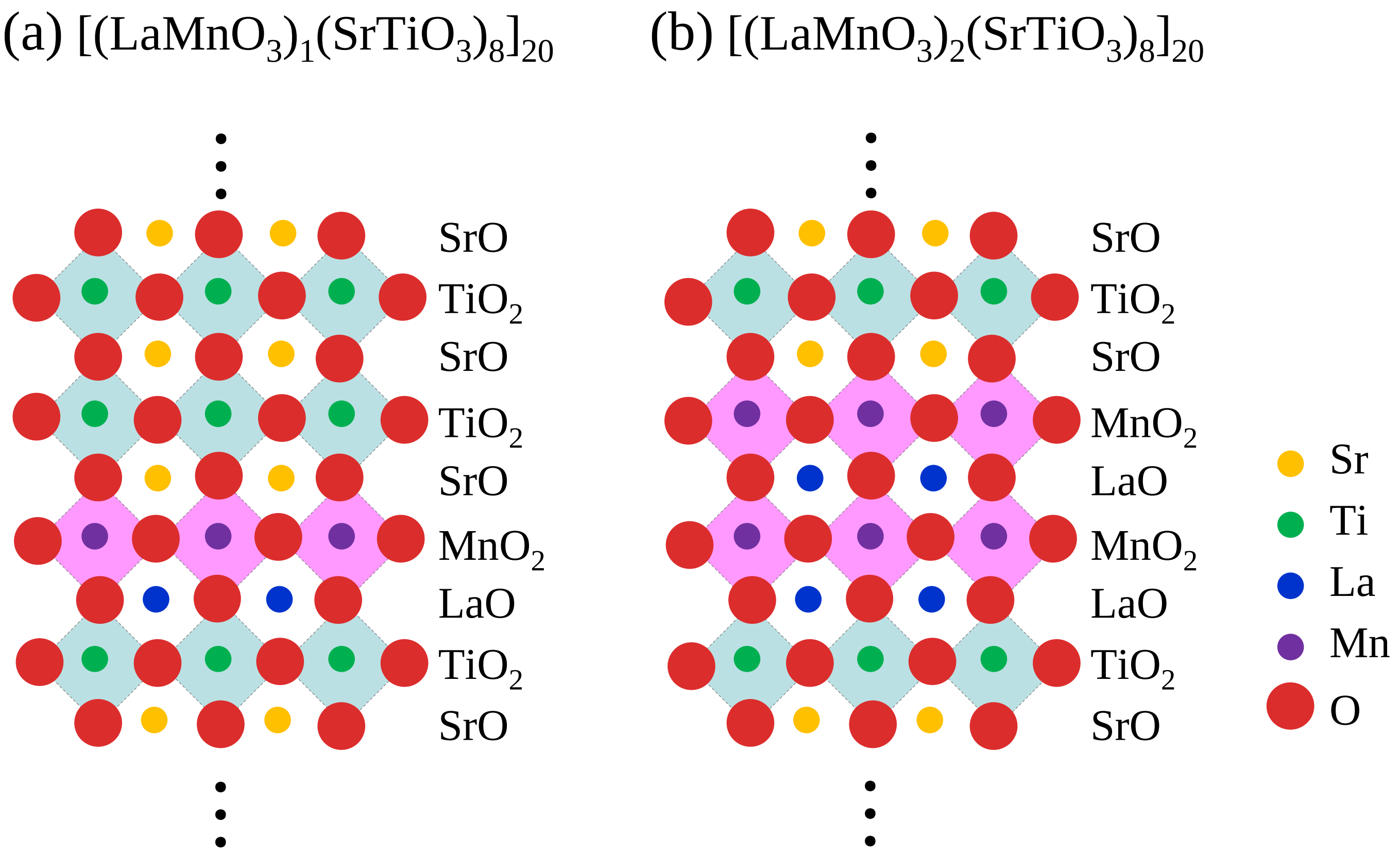

Fig. 2.
Choi *et al.*

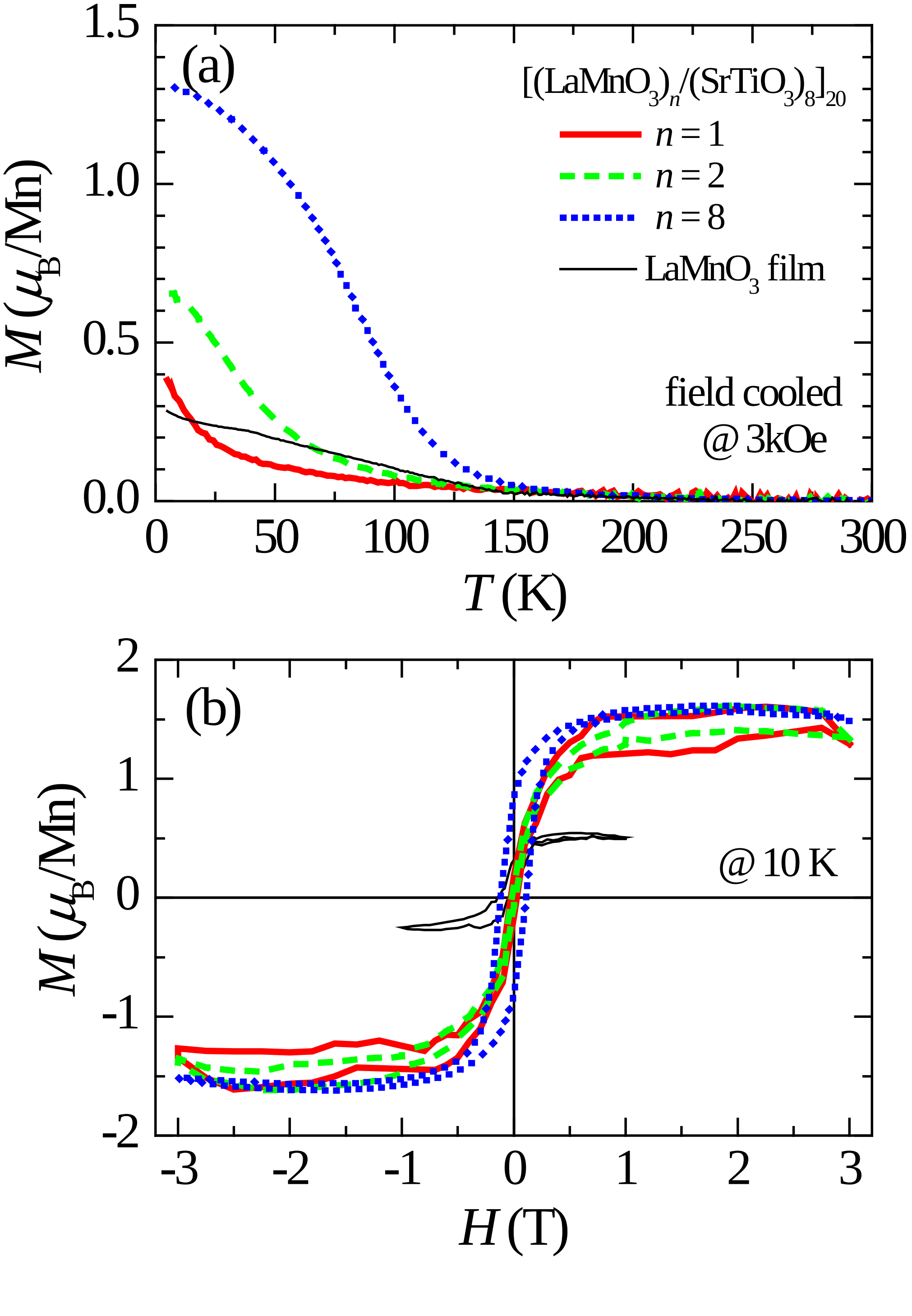

Fig. 3.
Choi *et al.*

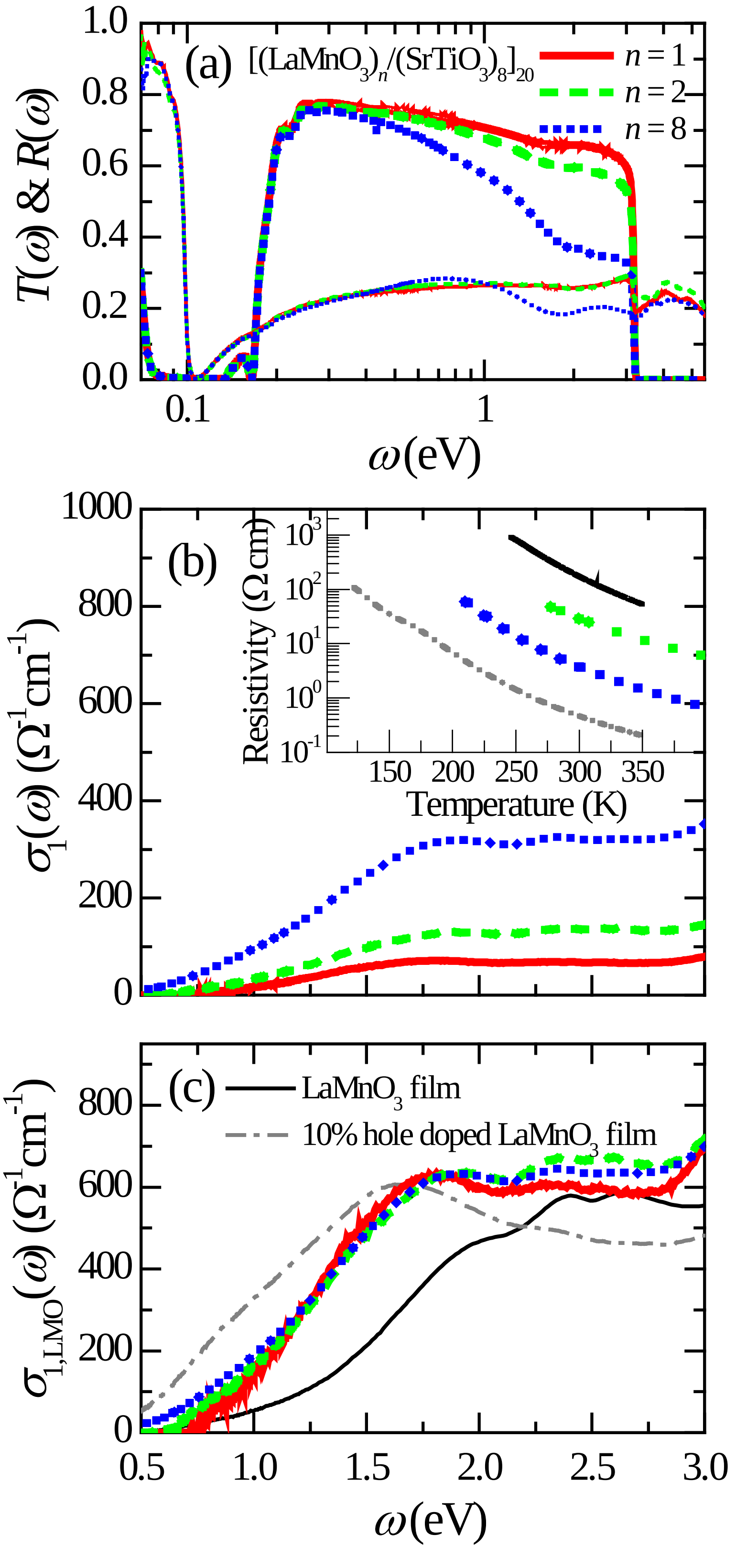

Fig. 4.
Choi *et al.*